# How to measure the local Dzyaloshinskii-Moriya Interaction in Skyrmion Thin-Film Multilayers

Mirko Baćani,[1] Miguel A. Marioni,[1,*] Johannes Schwenk,[1,2] Hans J. Hug[1,2]

[1] Empa, Swiss Federal Laboratories for Materials Science and Technology, CH-8600 Dübendorf, Switzerland

[2] Department of Physics, University of Basel, CH-4056 Basel, Switzerland

**ABSTRACT**

**The current-driven motion of skyrmions in MnSi and FeGe thinned single crystals could be initiated at current densities of the order of $10^6$A/m, five orders of magnitude smaller than for magnetic domain walls. The technologically crucial step of replicating these results in thin films has not been successful to-date, but the reasons are not clear. Elucidating them requires analyzing system characteristics at scales of few nm where the key Dzyaloshinskii-Moriya (DM) interactions vary, and doing so in near-application conditions, i.e. oxidation protected systems at room temperature. In this work's magnetic force microscopy (MFM) studies of magnetron-sputtered Ir/Co/Pt-multilayers we show skyrmions that are smaller than previously observed, are not circularly symmetric, and are pinned to 50-nm wide areas of 75% higher than average DM interaction. This finding matches our measurement of Co layer thickness inhomogeneity of the order of ±1.2 atomic monolayers per 0.6nm layer, and indicates that layer flatness must be controlled with greater precision to preclude skyrmion pinning.**

*Mini intro*

Ever since skyrmions were proposed in magnetic materials[1] and subsequently confirmed experimentally, first in MnSi[2] and FeCoSi,[3] it has been expected that due to topological stabilization[4] and sizes ranging between few and few hundred nm[5,6] skyrmions could find application in high density information storage,[7,8] and especially in low power spintronics devices and sensors,[9] particularly so after ultra-low critical currents densities for motion (in the range of $10^6$ A/m$^2$) were observed in MnSi[10] and FeGe[11] bulk non-centrosymmetric systems. Interestingly from the point of view of industrial fabrication[7], the DM interaction can also arise at interfaces in thin film multilayers,[12,13] such as Ir/Ni,[12] Ir/Co,[13,17] Pt/Ni,[12] Pt/Co,[13,17] Mn/W,[14] Ir/(FePd)[15], and Ta/Co/TaOx[16]. Skyrmions were recently observed e.g. in Ir/Co/Pt[17] and Pt/Co/MgO[18] based systems. However, the current density for the motion of these skyrmions remains high, at about $10^{11}$ A/m.[19]

To find out the reasons for the high threshold for motion we need to study the skyrmions at scales comparable to their width, where pinning likely arises, but multilayer systems pose a challenge for high resolution magnetic characterization. Spin polarized scanning tunneling microscopy (sp-STM) was able to image nm-sized skyrmions in a thin film[20] at cryogenic temperatures, mapping the spin orientation on the atomic scale for comparison with atomistic simulations. Yet to-date no room temperature sp-STM data of skyrmions has become available. Moreover, sp-STM cannot be used for oxidation protection-capped structures, or to assess DMI in application-oriented thicker multilayer systems. Lorentz transmission electron microscopy (LTEM), with which ultra-low current densities for skyrmion motion were demonstrated,[10,11] requires having electron-transparent samples, and thus excludes many cases of interest. Other methods used to assess magnetic domain wall chirality and skyrmion structures, such as x-ray magnetic circular dichroism with photoemission electron microscopy[18] (XMCD-PEEM) and spin-polarized low energy electron microscopy[12] (SPLEEM) have similar limitation concerning capping layers, whereas scanning transmission x-ray microscopy (STXM) requires x-ray transparent samples and currently cannot clearly distinguish 30 nm-wide skyrmions from domains or topographical features.

Here we use several MFM techniques to show that the skyrmions in a 5-repeat Ir/Co$_{0.6nm}$/Pt multilayer structure analogous to the 10-repeat one from Moreau-Luchaire *et al.*[17] are pinned at locations with a 75% larger than average DM interaction strength. Consistent with theoretical work[21] we associate these findings with magnetic layer thickness inhomogeneity, which we confirm by deconvolving the calibrated instrument point spread function from MFM measurements in saturating magnetic fields. The measurements show that the 0.6 nm thick Co layers vary on a 30 – 100 nm scale by up to ± 1.2 ± 0.2 atomic monolayers, with standard deviation from the mean of 0.3 atomic monolayers. Our observation that the skyrmions are not circularly symmetric, are pinned, and have return point memory can likewise be traced back to this inhomogeneity. An important difference to other MFM work on skyrmions[22] is operation at room temperature on thin films, using calibrated tips which requires improved sensitivity and tip-sample distance control. It allows us to distinguish skyrmions from more extended skyrmionic bubbles and model them accurately.

The starting point for this work is the confirmation that our Ir/Co$_{0.6nm}$/Pt-based multilayer (sample A; Figure 1a) has a large net DM interaction, compared with a control sample B comprising repeats of Pt$_{1nm}$/Co$_{0.6nm}$ (Figure 1c), which on account of its symmetric interfaces is not expected to have a net DM interaction (though small values cannot be excluded[23]). The as-grown and demagnetized states of sample A (Figures 1b) consist of maze patterns of magnetic domains. They are narrower than the as grown pattern from sample B (Figure 1d), which indicates a strong DM interaction.

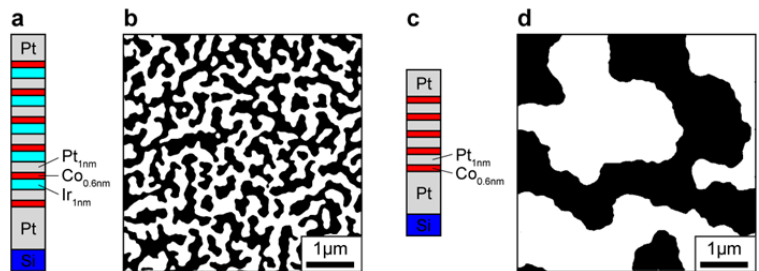

**Figure 1: domain pattern in the studied samples. a** Layout of the structure of sample A, in which the ferromagnetic Co layer is flanked on either side by different non-ferromagnetic materials, Ir and Pt. **b** Domain pattern of sample A (panel **a**) as grown. **c** Layout of the structure of sample B, in which the ferromagnetic Co layer is flanked symmetrically by Pt on either side. **d** Domain pattern of sample B (panel **c**) as grown.

The domain pattern contains information about the average DM interaction coefficient $D$, because it results from an equilibration of magnetostatic and wall energies. Through the latter the equilibrium domain configuration and the DM interaction are linked, a condition that can be exploited.

**Average Dzyaloshinskii-Moriya interaction**

For example, the width of regular stripe domains[24,25] that minimizes the total energy of the system (domain wall- plus magnetostatic energy) has been used[19] to estimate $D$. Two conceptual difficulties with this approach make it inadequate in our case. The first is that the experimental domains patterns do not closely resemble regular stripes, and any statistic of the domain pattern chosen to represent the regular stripe domain width is therefore a priori unreliable. The second difficulty, of particular severity in multilayer films such as ours, concerns the account of magnetostatic energy and DM interaction when the system is not a single homogeneous layer, as closed analytical expressions assume[19,24]. Other work therefore calculates the energy of the multilayer relying on micromagnetic simulations of domains that resemble the measured ones[17].

We use an alternative approach that circumvents the need for micromagnetic simulations, and allows us to rely solely on magnetometry data and the measured domain patterns obtained after demagnetization. Note that minimizing the total energy through variation of the width of domains in a regular stripe pattern[19] constitutes but one choice among many different variation paths in parameter space representing domain configurations. One could likewise select the spatial scale of a single arbitrary domain pattern for the variations. This approach is equivalent to varying the regular stripe domain width if and when the pattern of choice consists of regular stripes. More conveniently we can also choose a measured pattern, taking care it is close to the equilibrium. This avoids the problem of equating the experimental pattern to highly idealized stripes[19,24], and of selecting which measure of the experimental pattern best represents the width of stripe domains. Once a domain pattern has been selected the energy is calculated numerically. For a given $D$, which determines the domain wall energy,[26] we find the measured pattern's scale for which the total energy is a minimum. In general, the minimizing scale will depend on the $D$. At one specific $D$ value, however, the minimizing scale will coincide with the actual scale from the measurement. We take this $D$ value to represent the average value of the DM interaction coefficient best. Thus for domain patterns obtained after demagnetization in an oscillatory field with decaying amplitude (applied perpendicular to the film plane) the average DM coefficient is $D_{avg}$ = 1.97 ± 0.01 mJ/m$^2$ (Figure 2a; and Supporting Material).

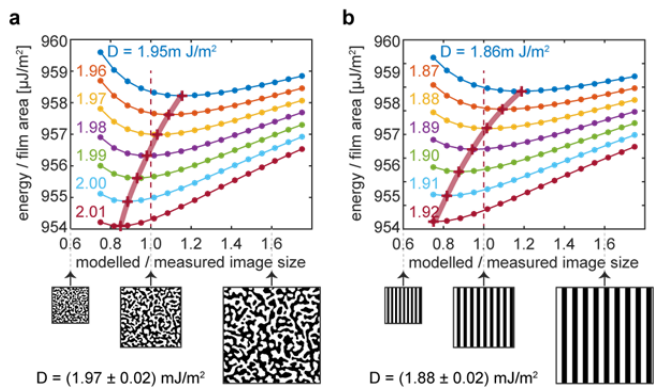

**Figure 2: Calculation of the average DMI coefficient $D$ from knowledge of an equilibrium pattern of domains. a** For a measured (perpendicular) demagnetized state pattern near the energy minimum, the

pattern scale is varied for a fixed parameter $D$ (colored labels), and the total energy of the system is computed, accounting for discrete Co layers. The pattern scales resulting in an energy minimum for different $D$ are connected with a broad red line as a guide. When the chosen parameter $D$ coincides with the film's actual average DMI coefficient the broad red line crosses the scale 1.0, i.e. the minimum energy is found for the actually measured scale (modelled/measured scale = 1.0). **b** Same as **a** except that a regular stripe pattern is assumed. Its width matches the 248 ± 26 nm average domain size measured from **a**.

This value is higher than in structures with more repetitions of the Ir/Co/Pt unit[17], which is likely related to the improved homogeneity attainable when fewer layers are used in the multilayer stack, as in our case. Conversely, our average $D$ is significantly lower than the value we would have expected based on theoretical predictions by Yang *et al.*[21], according to which $|D_{Pt/Co} + D_{Co/Ir}| \approx 3.26$ mJ/m² for a Pt/Co(3ML)/Ir system. Note that whereas in that work $D_{Co/Ir}$ = -0.34 mJ/m², experimental results of $D_{Co/Ir}$ = 0.55 meV/m² have been reported[12], so that a slightly lower value of $D \approx 2.36$ mJ/m² could also have been expected for our sample. The fact that we find $D_{avg}$ = 1.97 ± 0.01 mJ/m² could be an indication of interface roughness and intermixing in our multilayer system.

It is also noteworthy to point out that if we had modelled the system energy assuming regular stripe domains with a width given by the average domain size of 248 ± 26 nm obtained from the demagnetized domain pattern, we would have arrived at a significantly different effective $D$ of 1.88 ± 0.03 mJ/m² (cf. Figure 2b). Based on the estimate of $D_{avg}$ the DM interaction is too weak to result in a negative domain wall energy. Hence we do not expect to observe arrays of skyrmions forming spontaneously, but isolated skyrmions remain a possibility.[27]

**Room temperature skyrmions in thin film multilayers**

The characteristics of the room temperature skyrmions of our Ir/Co/Pt-based multilayers become apparent in Figure 3. After erasing the demagnetized domain pattern of Figure 1b at −128 mT and then lowering the field amplitude to −1.1 mT we observe skyrmions (Figure 3a) as small stable light spots with approximately circular geometry (up magnetization). As expected, no skyrmion-lattice but only a few isolated skyrmions exist in the 4.5µm x 4.5µm area of Figure 3a. In increasing positive fields (+4 mT in Figure 3b, and +8.5 mT in Figures 3c and d), two of the skyrmions visible in Figure 3a burst (strip out) into larger domains (as indicated e.g. by the yellow arrows in Figures 3b and c), and new spots of reversed magnetization with a non-circular geometry appear (red contours in Figures 3d, e, and f). These larger reversal domains are reminiscent of bubble domains characterized by a larger area of constant up magnetization at their centers.

Overall, these measurements reveal that skyrmions in our system are isolated, burst and become small non-circular domains in small positive fields (i.e. parallel to their core), and coexist with slightly larger reversal domains with non-circular geometry in stronger positive fields (see Figure 3e and 3f). The data also show that skyrmionic bubbles (areas in 3b to 3f enclosed by red ellipses) with larger diameters and non-circular boundaries can coexist with the smaller skyrmions. Additional data taken at a different location on the same film in small negative fields following saturation at −128mT confirm that the skyrmions repeatedly appear at the same locations. Skyrmions are thus pinned at specific locations with physical properties different from those of most of the sample.

For detailed analysis we exclude those bright spots in Figures 3a-3f which on account of their distinctly irregular shape or larger size are likely beyond the bursting point, and thus inadequately analyzed in terms of skyrmion models . We acquire MFM images with 4.88 nm pixel width, shown in Figures 3g–3l. For example, Figures 3g and 3j show two groups of two skyrmions each. The upper skyrmion in Figure 3g has stronger MFM contrast and an almost circular geometry, whereas the lower one has a slightly elliptical appearance. Applying a weak field antiparallel to the core of the skyrmions erases them (the lower one at B = -2.9mT, Figure 3h, and the upper one at B = −3.5mT, Figure 3i), but both skyrmions re-appear if the field is set back to −1.1 mT. Two further skyrmions with a slightly elliptical shape are shown in Figure 3j. One of those is erased by a small field of −2.1 mT, whereas the other one remains stable up to a field of −18.5 mT (Figure 3k), becoming almost circular at this field. Figure 3l shows the MFM signal obtained at −80.4 mT after erasing the upper skyrmion of Figure 3k. The fact that the skyrmions strip out at different fields is a further indication that these systems are not adequately described in terms of average quantities, for which that behavior ought to be the same.

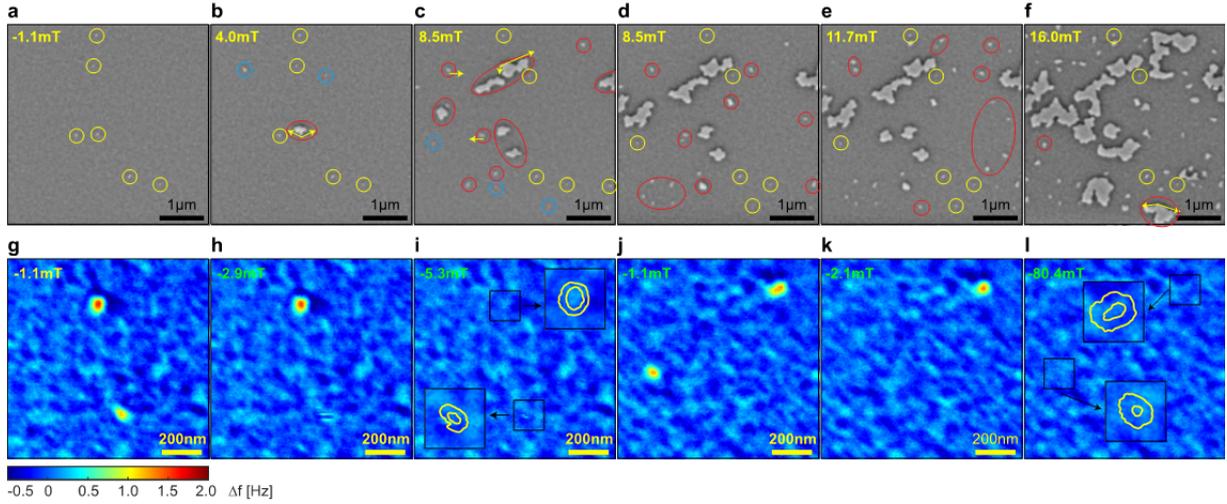

**Figure 3: Room temperature skyrmions in thin film multilayers. a-f** Sequence of images of a same area of sample A in increasing applied fields following negative saturation. **g-i** High resolution image sequence of a one area of sample A containing two skyrmions up to saturation (**i**). The inset panels zoom in (2x) onto the position of the skyrmions (150nm x 150nm squares) in relation to the background at saturation. **j-l** Second set of high resolution images.

The above observations imply that the values of the DM interaction coefficient $D$ are dependent on position. Conversely, a detailed account of the magnetic structure's characteristics should provide, in principle, these position-dependent values. However, in practice, the insufficient resolution of the measurement method can be an impediment. That fact limited the information on local $D$ that could be extracted from STXM data in prior work [17]. In this work we exploit the greater signal to noise ratio and spatial resolution of the MFM measurement to assess the position dependence of $D$. But we need to account for the fact that imaging at the limits of resolution, the details of the instrument's transfer function are important.

**Tip transfer function of the MFM / quantitative evaluation**

To see this, consider that the image formation in an MFM system can be described with transfer functions in 2D reciprocal space. From a computational point of view, and in terms of the 2D reciprocal space vector $\mathbf{k} = (k_x, k_y)$ and the spatial coordinate $z$, our instrument's transfer function $ICF(\mathbf{k}, z)$ enables us to relate the frequency shift map $\Delta f(\mathbf{k}, z)$ that our instrument reports to the sampled stray field (gradient) map $H_z(k, z)$:

$$\Delta f(\mathbf{k}, z) = ICF(\mathbf{k}, z) \frac{dH_z}{dz}(\mathbf{k}, z) \qquad (1)$$

Knowledge of $ICF(\mathbf{k}, z)$ therefore implies the capability of comparing experimental ($\Delta f$)- and model results ($\frac{dH_z}{dz}$, or $H_z$) in terms of their patterns but also in terms of their magnitude. It is the basis for the analysis of local variations of the measured skyrmions.

$ICF(\mathbf{k}, z)$ is a system-specific a priori unknown function, that is, it depends on cantilever stiffness, resonance frequency, oscillation amplitude and canting, tip-sample distance, and magnetic tip in a way that needs to be calibrated. The distance (and thickness-) loss factors[28,29] that describe the stray fields of sinusoidally varying magnetic charge distributions confer it a strong dependence on $\mathbf{k}$ (see Figures 4a and b). Determining it for our system constitutes the calibration process[30]. It requires knowledge of Δf and corresponding $H_z$ for at least one case. We use sample B, which has perpendicular anisotropy and uniform magnetization through-thickness, and has adequately sized domains of 437±64 nm in the in-plane demagnetized state. Note that a strong signal to noise ratio is needed in $\Delta f(\mathbf{k}, z)$ for an accurate calculation of $ICF(\mathbf{k}, z)$, which requires us to measure as close to the sample as possible, yet without altering the tip through contact. We accomplish this with a capacitive tip-sample distance control running on the second resonance of the cantilever.[33] In Figure 4c we show a one dimensional representation of the result for our instrument's transfer function, more specifically the radially averaged amplitude.

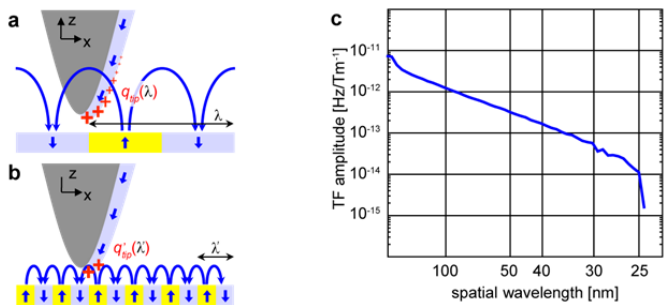

**Figure 4: Tip transfer function and its application. a** Schematic of the MFM tip in the stray field of wide domains. An extended part of the magnetic "charges" of the tip interact with the field. **b** The stray field of narrow domains has weaker intensity at the position of the tip. Only the fraction of tip magnetic "charges" closest to the sample participates in the interaction. **c** Dependence of the tip transfer function amplitude on the spatial wavelength of the domain pattern obtained from the tip calibration.

Before making use of the calibration thus obtained to assess in detail the local skyrmion characteristics, we turn our attention to an important observation about the background in the images,[17,19] which our improved instrument sensitivity allows making.

**Information contained in the magnetic background signal**

The contrast of the 'background', that is, the signal acquired far from the domain boundaries and skyrmions, is not uniform as one might expect at saturation. Here it amounts to an MFM contrast of ± 0.5 Hz (see Figure 5a showing MFM data acquired between the skyrmions shown in Figure 3). These contrast variations occur with length scales of 100 nm and below. This scale is comparable to the skyrmions' width. From the relation between inhomogeneity in the domain walls' energy landscape and its effect on pinning[31], we anticipate that if the background is magnetic it could have large influence on the skyrmion stability. It is therefore important to establish the origin of this background inhomogeneity.

Because the background remains constant across all field levels we know that its structure is not caused by instrument noise. Considering potentially relevant interactions we conclude it could arise at shallow topographical features (roughness) by van der Waals interaction between tip and sample or result from an inhomogeneous sample magnetic moment areal density, the stray field of which would interact with the tip magnetization. Note that in the former case the interaction is independent of the magnetization, whereas in the latter it changes sign with the relative sign of tip and sample magnetization. This distinguishing feature allows establishing that the background is predominately of magnetic origin (see Supporting Material).

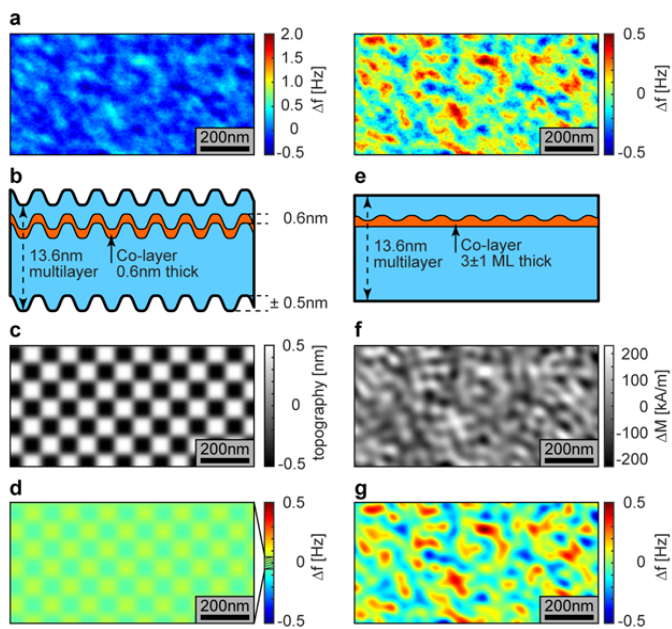

**Figure 5**: **Analysis of the magnetic background: a** High resolution image of an area of sample A in saturation. The left and right panels depict the same data with different contrast scales for clarity, with the left panel

corresponding to Figures 3g – 3l. **b** Schematic of a film with uniform 0.6 nm thickness and ± 0.5 nm roughness. **c** 2D topography representation corresponding to the schematic of **b**. **d** Calculated MFM image for the model system of **c**. The contrast amplitude is ± 0.05 Hz. **e** Schematic of a film with non-uniform thickness, with 3 ± 1 atomic monolayer thickness variations (0.6 ± 0.2nm). **f** Magnetization inhomogeneity obtained by deconvolution from the measured data **a** using the instrument calibration function.

Recall that the stray field of perfectly flat magnetic films of exactly uniform thickness and constant magnetization would vanish, and consider then the possibility that the magnetic background signal is caused by departures from this condition, i.e. roughness (Figure 5b).

In our measurements the average tip-sample distance is actively controlled using slow feedback parameters such that the tip follows the average tilt of the sample, but does not compensate for local variations of the sample topography, which amount to ± 0.5 nm (not shown). Using the calibrated $ICF$ we can calculate the MFM contrast for a saturated film that has a periodical roughness of ± 0.5 nm (Figure 5d). We thus find that the contrast can be at most ± 0.05 Hz, in effect ruling out the possibility of roughness as the cause for the contrast.

Consider next the possibility of thickness inhomogeneity. It can be described with varying degrees of complexity, but for definiteness we assume here that it affects all Co-layers identically. The roughness of the magnetic surfaces that is necessarily implied can be disregarded on account of the negligible stray field associated with it, as we have argued above. What remains is a flat layer of magnetization with inhomogeneous amplitude $\Delta M(k)$. It gives rise to a contrast that can be calculated by convolution of the calibrated $ICF(\boldsymbol{k}, z)$ and the stray field of each layer of average thickness $t_{Co}$ at a reference distance, as per Equation (1). In that equation, we can now express $H_z(\boldsymbol{k}, z)$ as

$$H_z(\boldsymbol{k}, z) = \frac{1}{2}\left[\sum_{i=1}^{6}\left(1 - e^{-k\,(i-1)(t_{Co}+t_{Pt}+t_{Ir})}\right)\right](1 - e^{-k\,t_{Co}})\Delta M(k)$$

(2)

where $t_{Pt}$ = 1nm and $t_{Ir}$ = 1nm are the (average) thickness of the Pt and Ir layers. $\Delta M(k)$ can now be found from the measured background patterns (Figure 5a, and also Figure 3i and l) by deconvolution. For the background contrast (Figure 5a) measured between the skyrmions shown in Figure 5a (and Figures 3i and 3l), we find a standard deviation of the magnetization of ± 64 kA/m and a maximum magnetization amplitude of ±230 kA/m (Figure 5e, and Figure 3i and 3l), the latter amounting to ± 1.2 monolayers of Co. Note that according to the calculations of Yang et al.[21] we would expect to observe a local DMI above 4 mJ/m$^2$ for a perfect Pt/Co(2ML) structure, and lower values for our multilayers with slightly diffuse or rough interfaces.

**Skyrmions and local values of DMI**

Previous work was able to estimate a large net $D$ in Ir/Co/Pt based multilayers by matching simulated magnetization profiles from skyrmions to the FWHM of the measured contrast spots[17]. The small skyrmions size, estimated to range between 30 and 90 nm, precluded a closer examination of variability in $D$ with these measurements. Similarly, chiral domain walls with strong $D$ were found in Pt/Co/MgO systems[18], but the 130 nm wide chiral bubble domains stabilized through geometrical confinement could not be used to estimate local values of $D$.

To derive local values of $D$ from the quantitative evaluation of high resolution MFM measurements of skyrmions, we model skyrmions following the original work from Bogdanov et al.[32] in which the reduced magnetization profile $\boldsymbol{M}$ of the skyrmion is required to minimize the reduced energy functional

$$w = \sum_i \left[ \left(\frac{\partial \boldsymbol{M}}{\partial \tilde{x}_i}\right)^2 - \tilde{\beta} M_z^2 - \boldsymbol{M} \cdot \boldsymbol{H}_{ext} - \frac{1}{2}\boldsymbol{M} \cdot \boldsymbol{H}_d + \omega_D \right] \qquad (3)$$

where $\boldsymbol{H}_{ext}$ and $\boldsymbol{H}_d$ are the external- and demagnetization fields[32] reduced with $H_D \equiv D^2/A\,K$ ($D$ is the DMI coefficient, $A$ the exchange stiffness and $K$ the anisotropy energy), $\tilde{\beta} \equiv AK/D^2$ is the reduced anisotropy constant, and $\omega_D = -D\left[M_x \frac{\partial M_z}{\partial \tilde{x}} - M_z \frac{\partial M_x}{\partial \tilde{x}} + M_y \frac{\partial M_z}{\partial \tilde{y}} - M_z \frac{\partial M_y}{\partial \tilde{y}}\right]$ accounts for the DMI. The dependence of $\boldsymbol{H}_d$ from the magnetization distribution over the whole film volume complicates the calculation except for infinitely thin- ($\boldsymbol{H}_d = -\boldsymbol{M}$) or infinitely thick ($\boldsymbol{H}_d = 0$) media in which case it takes on a simple form that can be adsorbed in the anisotropy. In our present case that approximation would not be warranted. Instead we replace $\boldsymbol{H}_d = -\alpha^2 \boldsymbol{M}$ with a constant $\alpha$ chosen such that the numerical integration over the multilayer $V$ ensures

$$\frac{1}{2}\int_V \alpha^2 \boldsymbol{M}_{Co} \cdot \boldsymbol{M}_{Co}\, dv = -\frac{1}{2}\int_V \boldsymbol{M}_{loc.} \cdot \boldsymbol{H}_d\, dv, \qquad (4)$$

where $M_{Co}$ is the magnetization of a single Co layer containing a skyrmion formally taken to be uniform through thickness, $H_d$ is the local demagnetization or stray field, and $M_{loc}$ is the magnetization distribution of the stack of Co layers, i.e. it is equal to $M_{Co}$ inside the Co layers and is set to zero elsewhere in the film. With the replacement $H_d = -\alpha^2 M$ in Eq. (3) the corresponding Euler equation can readily be solved numerically (e.g. using Matlab). In our case we take the exchange stiffness to be $A$ = 16 pJ/m, use the average uniaxial anisotropy $K_u$ = 414 kJ/m$^3$ and magnetization of the Co layer $M_{Co}$ = 653.6 kA/m (obtained from VSM). An initial value for $\alpha = \sqrt{t_{Co}/(t_{Co} + t_{Pt} + t_{Ir})} \approx 0.522$ allows finding a first skyrmion profile, to be used in refining $\alpha$ according to Eq. 4 (see Figure 6a). Self-consistency is attained after one iteration. $\alpha$ varies slightly for different choices of $D$, as is apparent from the plot of Figure 6a.

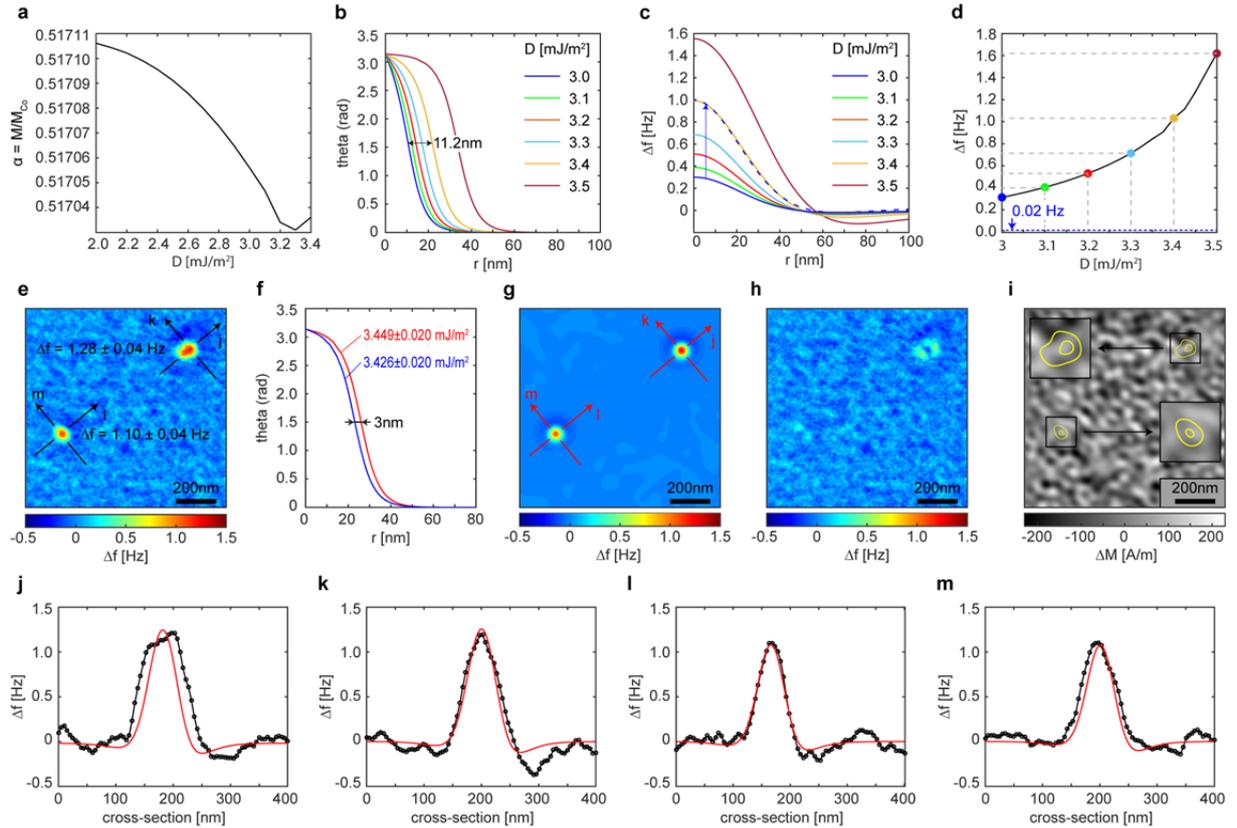

**Figure 6**: **skyrmion measured and simulated profiles.** **a** Dependence the factor α from $D$. **b** Calculated skyrmion magnetization profiles using the indicated values of $D$ and the corresponding factor α from **a**. **c** Simulated measurement profiles from **b** using the calibrated instrument transfer function, $ICF(\mathbf{k}, z)$. The dashed line shows what one would obtain if the transfer function's amplitude had been selected so that the simulation for $D = 3.0$ mJ/m$^2$ reached a peak contrast of 1 Hz, comparable to the simulation for $D = 3.4$ mJ/m$^2$. **d** Calculated measurement contrast at skyrmion centers as function of $D$ (using **c**) . **e** Measurement of a region of sample A containing two skyrmions at 1.14 mT. The background has been subtracted. Sections j-l are reproduced in the corresponding black trace of panels **j-l**. **f** Calculated skyrmion profiles for the values of D

corresponding to the maximum contrast of top and bottom skyrmions. **g** Calculated measurement contrast maps for the skyrmion profiles of **f**. The sections j-l are reproduced in the corresponding red trace of panels **j-l**. **h** Point by point map difference **e−g**. **i** Magnetization pattern which repeated on each Co layer reproduces the measured contrast in saturation, i.e. a measure of the local thickness' departure from the average 0.6 nm. **j-m** cross sections indicated in **e−g**.

In this way we obtain the skyrmion magnetization orientation profiles e.g. for 3 mJ/m$^2$ < $D$ < 3.5 mJ/m$^2$, which we show in Figure 6b. From them, we can directly produce a circularly symmetric magnetic charge pattern for each skyrmion (not shown) and then the corresponding $z$-component of the stray field (not shown) at the tip sample distance of 12 nm, used for the MFM measurements. This enables us to simulate the measured contrast by convolving the stray field map with the instrument calibration function, $ICF(\mathbf{k}, z)$, from where we can obtain the profiles in Figure 6c.

As in most magnetic imaging methods, with the exception of sp-STM, in MFM the effective probe-sample interaction is a weighted average over a characteristic volume. Because of that, narrow magnetic objects can appear wider than they actually are. This means that small magnetic objects of arbitrary profile could result in contrast peaks resembling skyrmions. A particularly striking example of this effect is given by scaling the MFM profile of skyrmion with D = 3.0 mJ/m$^2$ to the same contrast as that with D = 3.4 mJ/m$^2$ (dashed blue line in Figure 6c). Their apparent widths are roughly the same although the spin profile (Figure 6b) differ by 30 nm. It becomes clear that calibrating the tip- and instrument-specific $ICF(\mathbf{k}, z)$ is necessary to remove this scaling degree of freedom. Figure 6d shows the peak contrast from the skyrmion simulations in Figure 6c, and indicates that the maximum contrast is a proxy for $D$, provided the instrument calibration function is known. However, note that skyrmions with $D = 1.97$ mJ/m$^2$ would produce a contrast that is below noise level of our measurement (± 0.02 Hz, cf. dashed horizontal line in Figure 6d).

The experimental data in Figure 3j (See supporting material for a further example based on the data of Figure 3g) contains a magnetic background, as argued, which we remove prior to comparison with simulations, i.e. we pointwise subtract the map of Figure 3l (-80.4mT) from that of Figure 3j (-1.1mT). This results in Figure 6e out of which we extract the maximum contrast of the upper and lower skyrmion by averaging over several equivalent background subtracted images of the same area. The MFM contrast of the upper and lower skyrmion is

1.28 ± 0.04 and 1.10 ± 0.04 Hz respectively. With Figure 6d $D$ = 3.449 ± 0.020 mJ/m$^2$ and 3.426 ± 0.020 mJ/m$^2$ and for the upper and lower skyrmion respectively.

Note that the difference in skyrmion profile width is about 3 nm (Figure 6f). The complete validation of the fit comes from the detailed comparison of the simulated skyrmion maps. These we plot in Figure 6g. The sections across the skyrmions in images 6e and 6g can be seen in Figures 6j-6m. The black trace is the experimental result, highlighted in corresponding lines in Figure 6e, whereas the superimposed red traces are the simulation result from Figure 6g.

Figure 6h shows the pointwise difference between experiment (Figure 6e) and simulation (Figure 6g). By construction, the simulation fits very well the center contrast of the skyrmions, and an approximately 30 × 30 nm$^2$ wide region surrounding it, but the quality of the fit degrades away from the center in some cases. Specifically, the deterioration is minimal for the cross-sections k and l but it is observable for the cross-sections j and to a lesser extent m (Figures 6j to 6m). This indicates that the large values of $D$ are circumscribed to small areas. Furthermore, matching the width of the skyrmion along the j-directions leads to a large overestimate of the center contrast (not shown). Consequently $D$ must be changing along the j-direction (in particular) over length scales of the order of 30 nm which the isotropic model in Equation (3) cannot capture in detail. Importantly, the large inhomogeneity of $D$, which we showed to be as much as 75% larger than the average, is a cause for pinning, of the kind we would expect to raise the current density threshold for motion, and consistent with skyrmions not moving when a field is applied.

Figure 6i displays the calculated background magnetization variation (i.e. deconvolution of our $ICF$ from the measurement in saturation), which is a proxy for the average layer thickness, and superimposes contours of the skyrmions. The figures show that the skyrmions are invariably localized in areas of relatively low relative magnetic signal (bright contrast), i.e. thin Co layers, which the zoomed-in insets show. This is consistent with the previously calculated strong film thickness dependence of $D$[21]. Regions in which all Co layers have the same constant small thickness are less likely to occur in our sputtered thin film, and therefore areas of high $D$ are correspondingly sparse, consistent with our observations.

It is somewhat surprising that despite large significant differences between average and local values of $D$ the analyzed skyrmions have a rather narrow distribution of $D$. The observation may be explained with the sparseness of large-$D$ regions (and low thickness regions found by our analysis of the magnetic background contrast), and is connected with the observed absence of skyrmions with a weaker $D$ than for the four assessed skyrmions. To understand why, consider that when we reduce the applied field from negative saturation to −1.1mT the high $D$ skyrmions, particularly for $D > D_{\kappa=1} \approx 3.28$ mJ/m$^2$ (the skyrmion lattice threshold value) are the first to nucleate stably because their non-uniform magnetization lowers the energy (cf. the stability phase diagram in e.g. Kiselev et al.[27]). As anticipated for our measured average $D$ of 1.97mJ/m$^2$ isolated skyrmions can also exist for $D < D_{\kappa=1}$, but their nucleation would require overcoming a barrier from the non-uniform magnetization in this case, and would be delayed to larger fields parallel to the skyrmion core (reversed or positive fields). Furthermore, such low-$D$ skyrmions burst into large domains at relatively low reversed fields[27]. This is consistent with our experimental observation (c.f. Figure 3). When lowering the field from negative saturation and reversing it to +1.1 mT, therefore, skyrmions at film positions with a high $D$ nucleate first, and are then stabilized by surrounding areas of lower $D$, which hinder their expansion into bubble domains at sufficiently small reversed (positive) fields. As the reverse field is increased, skyrmions with lower $D$ can nucleate, but may immediately expand into a (reversed) bubble domain (see e.g. red circle in Figure 3b), or will be swallowed by an expanding reversed domains arising from bursted skyrmions with higher $D$.

**Conclusion**

We observed skyrmions corresponding to DM-stabilized high-curvature core magnetization profiles in sputtered thin films multilayers. The DM interaction at the location of the skyrmions can exceed the average value of 1.97 ± 0.02 mJ/m$^2$ by as much as 75% in our case, reaching 3.45 ± 0.04 mJ/m$^2$. At this level the skyrmion lattice should be stable, showing that the skyrmion lattice could exist in sputtered thin films with additive DMI, which is critical for applications. But the inhomogeneity in the DMI is of the order of the skyrmion size, so that it will strongly affect the shape of the skyrmions and lead to strong pinning, as we see. The underlying cause of the DMI inhomogeneity is the Co-thickness, revealed by the stray fields at saturation which are consistent with thickness variability in the Co layers up to

±1.2 monolayer and standard deviation of 0.3 nm. Because of the connection of $D$ with layer thickness the skyrmions are not circularly symmetric; more extensive theoretical studies of skyrmion profiles in inhomogeneous $D$ will benefit from comparison with these data. Overall, MFM offers a convenient way for studying the average and local $D$, and can even provide a measure of the subtle thickness variations in industrially relevant multilayer systems.

**Methods**

**Film structures and preparation.** We prepared two thin film magnetic multilayer structures by DC magnetron sputter deposition at room temperature, using Ar gas at $1.8\times10^{-3}$ mbar. The system's base pressure is $2\times10^{-8}$ mbar. Sample A is $Si^{(nat.)}\backslash Pt_{10nm}\backslash Co_{0.6nm}\backslash Pt_{1nm}[Ir_{1nm}\backslash Co_{0.6nm}\backslash Pt_{1nm}]_{x5}\backslash Pt_{3nm}$ where $Si^{(nat.)}$ stands for the Si (100) substrate with native oxide. Sample B is a control sample sputtered analogously to sample A but with symmetric interfaces, i.e. $Si^{(nat.)}\backslash Pt_{10nm}\backslash [Co_{0.6nm}\backslash Pt_{1nm}]_{x5}\backslash Pt_{3nm}$. compensated by the Co\Pt interface's at the other side. This symmetry is broken in sample A and a net DMI can arise. A first manifestation of this fact is the noticeably different size (> 1000 nm in sample A and (246 ± 40) nm in sample S of the magnetic domains in the as-grown state observed by MFM in either case, shown in Figs. 1B,H.

**Magnetic characterization:** For the macroscopic magnetic characterization we used a vibrating sample magnetometer (Quantum Design PPMS). We obtain an anisotropy $K_u$ = 414 kJ/m$^3$, magnetization of the Co layers of 653.6 kA/m, resulting in an effective anisotropy of $K_{eff}$ = 146 kJ/m$^3$. For the microscopic characterization we used a room temperature magnetic force microscope operated in vacuum ($10^{-6}$ mbar) for better sensitivity and a Zurich Instruments Lockin/PLL system. We use a SS-ISC cantilever from team Nanotec GmbH (< 5 nm tip radius) coated in-house. For the measurement we operate a dual mode non-contact method with capacitive control[33] of the tip-sample distance to 12.0 ± 0.5 nm.

**Supplementary Information** is linked to the online version of the paper at […] .

**Acknowledgements** This work was supported by Empa. J. S. was supported by SNF (Sinergia 'Understanding nanofriction and dissipation across phase transitions'). We gratefully acknowledge CCMX for support through the "Quantitative Magnetic Force Microscopy platform (qMFM)", in which the Matlab analysis platform (http://qmfm.empa.ch) was programmed by Zoe Goey, Sven Hirsch and Gabor Székely. We also would like to thank Sara Romer and Alexandre Guiller for the film fabrication and magnetometry, and acknowledge Sasa Vranjkovic for the design and construction of instrument components.

**Author contributions** M.A.M., H.J.H. co-designed the materials and experiments. M.B. and H.J.H carried out MFM measurements. H.J.H., M.A.M. and J.S. analyzed the data. M.A.M. and H.J.H. wrote the manuscript. All authors discussed the results and the text of the manuscript.

**Author information** The authors declare no competing financial interests. Correspondence should be addressed to M.A.M. (Miguel.Marioni@empa.ch).

**References**

1. Bogdanov, A. N. & Yablonskii, D. A. Thermodynamically stable' vortices' in magnetically ordered crystals. The mixed state of magnets. *Sov. Phys. JETP* **68**, 101-103 (1989).
2. Mühlbauer, S. et al. Skyrmion lattice in a chiral magnet. *Science* **323**, 915–919 (2009).
3. Yu, X. Z. et al. Real-space observation of a two-dimensional skyrmion crystal. *Nature* **465**, 901–904 (2010).
4. Oike, H. et al. Interplay between topological and thermodynamic stability in a metastable magnetic skyrmion lattice. *Nature Phys* **12**, 62–66 (2016).
5. Nagaosa, N. & Tokura, Y. Topological properties and dynamics of magnetic skyrmions. *Nature Nano.* **8**, 899–911 (2013).
6. Wiesendanger, R. Nanoscale magnetic skyrmions in metallic films and multilayers: a new twist for spintronics. *Nature Rev. Mater.* **1**, 16044 (2016).
7. Fert, A., Cros, V. & Sampaio, J. Skyrmions on the track. *Nature Nano* **8**, 152–156 (2013).
8. Tomasello, R. et al. A strategy for the design of skyrmion racetrack memories. *Sci. Rep.* **4**, (2014).
9. Hoffmann, A. & Bader, S. D. Opportunities at the Frontiers of Spintronics. *Phys. Rev. Applied* **4**, 047001 (2015).
10. Jonietz, F. et al. Spin Transfer Torques in MnSi at Ultralow Current Densities. *Science* **330**, 1648–1651 (2010).
11. Yu, X. Z. et al. Skyrmion flow near room temperature in an ultralow current density. *Nature Commun* **3**, 988 (2012).
12. Chen, G. et al. Tailoring the chirality of magnetic domain walls by interface engineering. *Nature Commun.* **4**, (2013).
13. Chen, G., N'Diaye, A. T., Wu, Y. & Schmid, A. K. Ternary superlattice boosting interface-stabilized magnetic chirality. *Appl. Phys. Lett.* **106**, 062402 (2015).
14. Ferriani, P. et al. Atomic-Scale Spin Spiral with a Unique Rotational Sense: Mn Monolayer on W(001). *Phys. Rev. Lett.* **101**, 027201 (2008).
15. Romming, N. et al. Writing and Deleting Single Magnetic Skyrmions. *Science* **341**, 636–639 (2013).
16. Jiang, W. et al. Blowing magnetic skyrmion bubbles. *Science* **349**, 283–286 (2015).
17. Moreau-Luchaire, C. et al. Additive interfacial chiral interaction in multilayers for stabilization of small individual skyrmions at room temperature. *Nature Nanotech.* **11**, 444–448 (2016).
18. Boulle, O. *et al.* Room-temperature chiral magnetic skyrmions in ultrathin magnetic nanostructures. *Nature Nanotech.* **11**, 449–454 (2016).
19. Woo, S. et al. Observation of room-temperature magnetic skyrmions and their current-driven dynamics in ultrathin metallic ferromagnets. *Nature Mater* **15**, 501–506 (2016).
20. Heinze, S. et al. Spontaneous atomic-scale magnetic skyrmion lattice in two dimensions. *Nature Phys.* **7**, 713–718 (2011).
21. Yang, H., Thiaville, A., Rohart, S., Fert, A. & Chshiev, M. Anatomy of Dzyaloshinskii-Moriya Interaction at Co/Pt Interfaces. *Phys. Rev. Lett.* **115**, 267210 (2015).
22. Milde, P. et al. Unwinding of a Skyrmion Lattice by Magnetic Monopoles. *Science* **340**, 1076–1080 (2013).
23. Hrabec, A. et al. Measuring and tailoring the Dzyaloshinskii-Moriya interaction in perpendicularly magnetized thin films. *Phys. Rev. B* **90**, 020402 (2014).
24. Málek, Z. & Kamberský, V. On the theory of the domain structure of thin films of magnetically uniaxial materials. *Czech J Phys* **8**, 416–421 (1958).
25. Millev, Y. Bose-Einstein integrals and domain morphology in ultrathin ferromagnetic films with perpendicular magnetization. *J. Phys: Condens. Matter* **8**, 3671 (1996).

26. Rohart, S. & Thiaville, A. Skyrmion confinement in ultrathin film nanostructures in the presence of Dzyaloshinskii-Moriya interaction. *Phys. Rev. B* **88**, 184422 (2013).
27. Kiselev, N. S., Bogdanov, A. N., Schäfer, R. & Rössler, U. K. Chiral skyrmions in thin magnetic films: new objects for magnetic storage technologies? *J. Phys. D: Appl. Phys*. **44**, 392001 (2011).
28. Bertram, H. N. Theory of Magnetic Recording. (Cambridge University Press, 1994).
29. Camras. Magnetic Recording Handbook. (Springer Science & Business Media, 2012).
30. van Schendel, P. J. A., Hug, H. J., Stiefel, B., Martin, S. & Güntherodt, H.-J. A method for the calibration of magnetic force microscopy tips. *J. Appl. Phys*. **88**, 435–445 (2000).
31. Paul, D. I. Extended theory of the coercive force due to domain wall pinning. *J. Appl. Phys.* **53**, 2362–2364 (1982).
32. Bogdanov, A. & Hubert, A. Thermodynamically stable magnetic vortex states in magnetic crystals. *J. Magn. Magn. Mater.* **138**, 255–269 (1994).
33. Schwenk, J. et al. Bimodal magnetic force microscopy with capacitive tip-sample distance control. *Appl. Phys. Lett*. **107**, 132407 (2015).